\documentclass[twocolumn,nofootinbib,
showpacs,aps,floatfix]{revtex4}
\usepackage{bm}
\usepackage{graphicx}
\def\la{\langle}
\def\ra{\rangle}
\def\beq{\begin{equation}}
\def\eeq{\end{equation}}
\def\beqa{\begin{eqnarray}}
\def\eeqa{\end{eqnarray}}
\newcommand {\fexp} [1] {\exp \left( #1 \right)}
\newcommand {\fabsq}[1] {\left| #1 \right|^2}
\newcommand {\fabs}[1] {\left| #1 \right|}
\newcommand {\Msi}{\times 10^6/\mbox{s}}

\newcommand {\mum}{\, \mu \mbox{m}}

\begin{document}
\title{Atom diode: Variants, stability, limits, and adiabatic  
interpretation}
\author{A. Ruschhaupt}
\email[Email address: ]{wtxruxxa@lg.ehu.es}
\affiliation{Departamento de Qu\'\i mica-F\'\i sica, Universidad del
Pa\'\i s Vasco, Apdo. 644, 48080 Bilbao, Spain}
\author{J. G. Muga}
\email[Email address: ]{jg.muga@ehu.es}
\affiliation{Departamento de Qu\'\i mica-F\'\i sica, Universidad del
Pa\'\i s Vasco, Apdo. 644, 48080 Bilbao, Spain}

\begin{abstract}
We examine and explain the stability properties of the ``atom diode'', 
a laser device that lets the ground state
atom pass in one direction but not 
in the opposite direction. The diodic behavior and the variants 
that result by using different laser configurations 
may be understood with an adiabatic 
approximation. The conditions to break down the approximation, which imply 
also the  
diode failure,  
are analyzed.

\end{abstract}
\maketitle

\section{Introduction}

In a previous paper \cite{ruschhaupt_2004_diode} we proposed simple
models for an ``atom diode'', a laser device 
that lets the neutral atom in its ground state pass in one direction 
(conventionally from left to right) but not 
in the opposite direction for a range of incident velocities.   
A diode is a very basic control element in a circuit, so many applications may be  
envisioned 
to trap or cool atoms, or to build logic gates 
for quantum information processing in atom chips or other setups.  
Similar ideas have been developed independently by Raizen and
coworkers \cite{raizen.2005,dudarev.2005}.
While their work has emphasized phase space compression, 
we looked for the laser interactions leading to the most effective diode. 
This lead us to consider first STIRAP \cite{STIRAP} 
(stimulated Raman adiabatic passage)
transitions and three level atoms, although we also proposed
schemes for two-level atoms.  
In this paper we continue the investigation on the atom diode, 
concentrating on its two-level version, by examining 
its stability with respect to parameter changes, and also several variants,
including in particular the ones discussed in 
\cite{ruschhaupt_2004_diode} and \cite{raizen.2005}.
We shall see that the behaviour of the diode, its properties, 
and its working parameter domain can be
understood and quantified with the aid of an
adiabatic basis (equivalently, partially 
dressed states) obtained by diagonalizing the effective interaction potential.
  
We restrict the atomic motion, similarly to \cite{ruschhaupt_2004_diode}, to  
one dimension. This occurs when the atom travels  
in waveguides formed by optical fields   
\cite{schneble.2003}, or by electric or magnetic interactions due 
to charged or current-carrying structures \cite{folman.2002}.
It can be also a good approximation in free space 
for atomic packets which are broad in the laser direction, perpendicular to 
the incident atomic direction 
\cite{HHM05}. Three dimensional effects 
should not imply a dramatic disturbance, in any case, as we shall analyze 
elsewhere.  
%
\begin{figure}[t]
\begin{center}
\includegraphics[angle=0,width=0.6\linewidth]{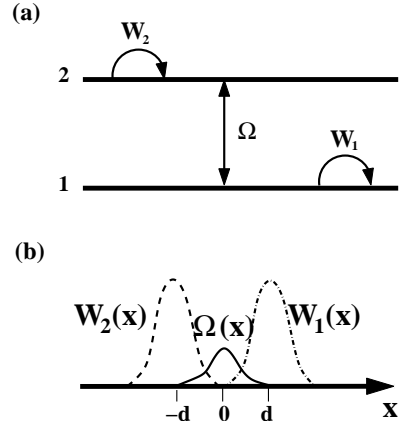}
\end{center}
\caption{\label{fig1}(a) Schematic action of the different lasers 
on the atom levels and 
(b) location of the different laser potentials.}
\end{figure}
%

The basic setting can be seen in Fig. \ref{fig1}, and consists of 
three, partially overlapping laser fields: two of them are
state-selective mirror lasers blocking the 
excited ($|2\rangle$) and ground  ($|1\rangle$)
states on the left and right, respectively
of a central pumping laser on resonance with the atomic transition.  
They are all assumed to be traveling waves 
perpendicular to the atomic motion 
direction. The corresponding effective, time-independent, interaction-picture 
Hamiltonian for the two-level atom
may be written, using     
$|1\rangle \equiv {1 \choose 0}$ and $|2\rangle \equiv {0 \choose 1}$, 
as  
\begin{eqnarray}
\bm{H} = \frac{\hat{p}_x^2}{2m} + 
\underbrace{\frac{\hbar}{2} \left(\begin{array}{cc}
W_1(x) & \Omega (x)\\
\Omega (x) & W_2(x)
\end{array}\right)}_{\bm{M}(x)},
\label{ham2}
\end{eqnarray}
where $\Omega(x)$ is the Rabi frequency for the resonant transition 
and the 
effective reflecting potentials are $W_1(x)\hbar/2$ 
and $W_2(x)\hbar/2$. 
$\hat{p}_x = -i\hbar \frac{\partial}{\partial x}$ is the momentum operator
and $m$ is the mass (corresponding to Neon in all numerical examples).

Spontaneous decay is neglected here
for simplicity, but it could be 
incorporated following  
\cite{ruschhaupt_2004_diode}. 
It  implies  both perturbing and
beneficial effects for unidirectional transmission. 
Notice that in the ideal diode operation the ground state atom must be excited  
during its left-to-right crossing of the device.  
In principle, excited atoms could cross 
the diode ``backwards'', i.e., from right to left, but  
an irreversible decay from the excited state to the ground
state would block any backward motion
\cite{ruschhaupt_2004_diode}.

The behaviour of this device is quantified by the scattering 
transmission and reflection amplitudes 
for left (l) and right (r) incidence.  
Using $\alpha$ and $\beta$ to denote the 
channels, $\alpha=1,2$, $\beta=1,2$,  
let us denote by $R^l_{\beta\alpha} (v)$ ($R^r_{\beta\alpha} (v)$)
the scattering amplitudes for incidence with modulus of velocity $v>0$
from the left (right) in channel $\alpha$, and reflection in channel $\beta$.  
Similarly we denote 
by $T^l_{\beta\alpha} (v)$ ($T^r_{\beta\alpha} (v)$) the scattering 
amplitude for incidence in channel $\alpha$ from
the left (right) and transmission in channel $\beta$.

For some figures, it will be preferable to use an alternative notation
in which the information of the superscript ($l/r$) is contained instead 
in the sign of the velocity argument $w$,
positive for left incidence and negative otherwise,  
\beqa  
R_{\beta\alpha}(w):=
\left\{
\begin{array}{ll}
R^l_{\beta\alpha}(\fabs{w}),&  {\rm{if}}\; w>0
\\
R^r_{\beta\alpha}(\fabs{w}),& {\rm{if}}\; w<0
\end{array}
\right.
\nonumber\\
T_{\beta\alpha}(w):=
\left\{
\begin{array}{ll}
T^l_{\beta\alpha}(\fabs{w}),&  {\rm{if}}\; w>0
\\
T^r_{\beta\alpha}(\fabs{w}),& {\rm{if}}\; w<0
\end{array}  
\right.
\nonumber
\eeqa
The ideal diode configuration must be such that 
\beqa
\fabsq{T_{21}^l (v)}\approx \fabsq{R_{11}^r (v)}  \approx 1,
\label{di1}
\\ 
\fabsq{R_{\beta 1}^l (v)} \approx \fabsq{T_{\beta 1}^r (v)} \approx
\fabsq{R_{21}^r (v)} \approx \fabsq{T_{11}^l (v)}
\approx 0,
\label{di2}
\eeqa
with $\beta=1,2$, 
in an interval $v_{min}<v<v_{max}$ of the modulus of the velocity.
In words, there must be full transmission for left incidence 
and full reflection for right incidence in the ground state.    
This was achieved in \cite{ruschhaupt_2004_diode} with 
a particular choice of the potential in which $\Omega(x)$, $W_1(x)$, 
and $W_2(x)$ 
were related to two partially overlapping functions $f_1(x)$, $f_2(x)$.
However, other forms are also possible, so we shall deal here with the  
more general structure 
of Eq. (\ref{ham2}). We shall use Gaussian laser profiles 
\begin{eqnarray*}
&W_1 (x) = \hat{W}_1 \;\Pi(x,d),\quad W_2 (x) =  \hat{W}_2 \;\Pi(x,-d),&\\
&\Omega (x) = \hat{\Omega}\; \Pi (x,0),&
\end{eqnarray*}
where 
$$
\Pi (x,x_0)=\exp[-(x-x_0)^2/(2\Delta x^2)]
$$
and $\Delta x = 15 \mum$.

In section \ref{s2} we shall examine the stability and limits of the ``diodic''
behavior  while the variants of the atom diode 
are
presented in section \ref{s3}. They are explained in 
section \ref{s4} with the aid of an adiabatic basis and 
approximation. 
The paper ends with a summary and an appendix on the adiabaticity 
criterion. 
\section{``Diodic'' behaviour and its limits\label{s2}}
%
%
%
%
%
%
%
%
\begin{figure}[t]
\begin{center}
\includegraphics[angle=-90,width=0.95\linewidth]{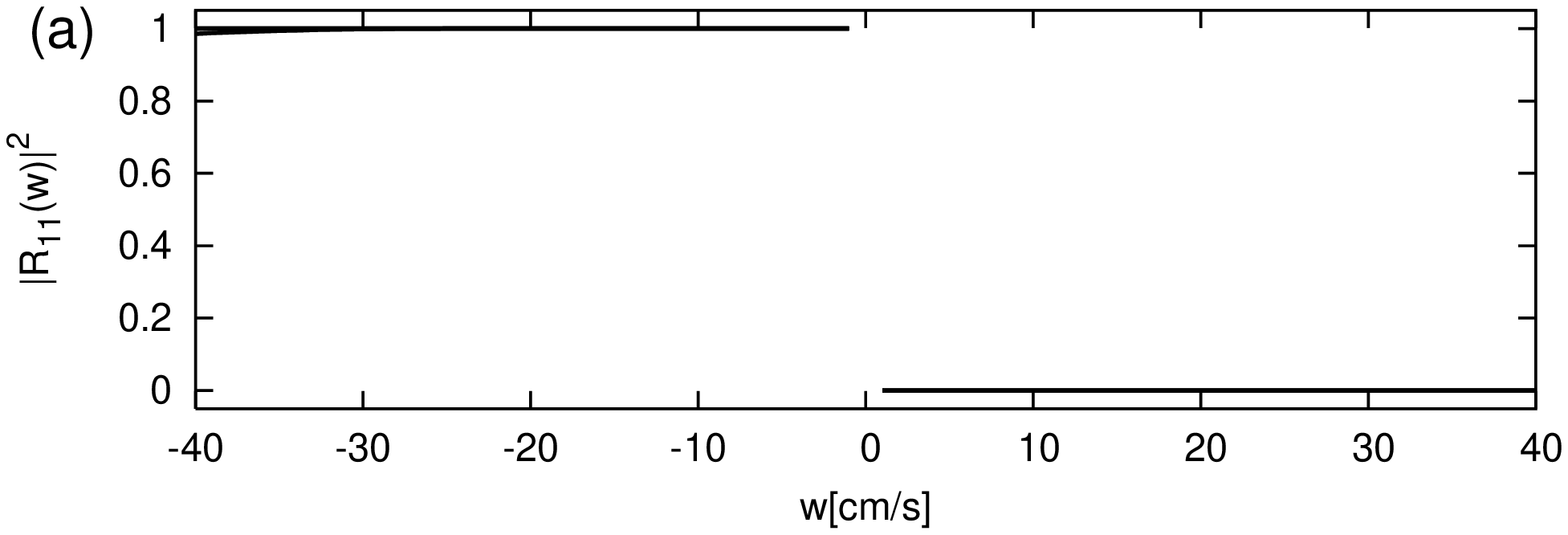}

\includegraphics[angle=-90,width=0.95\linewidth]{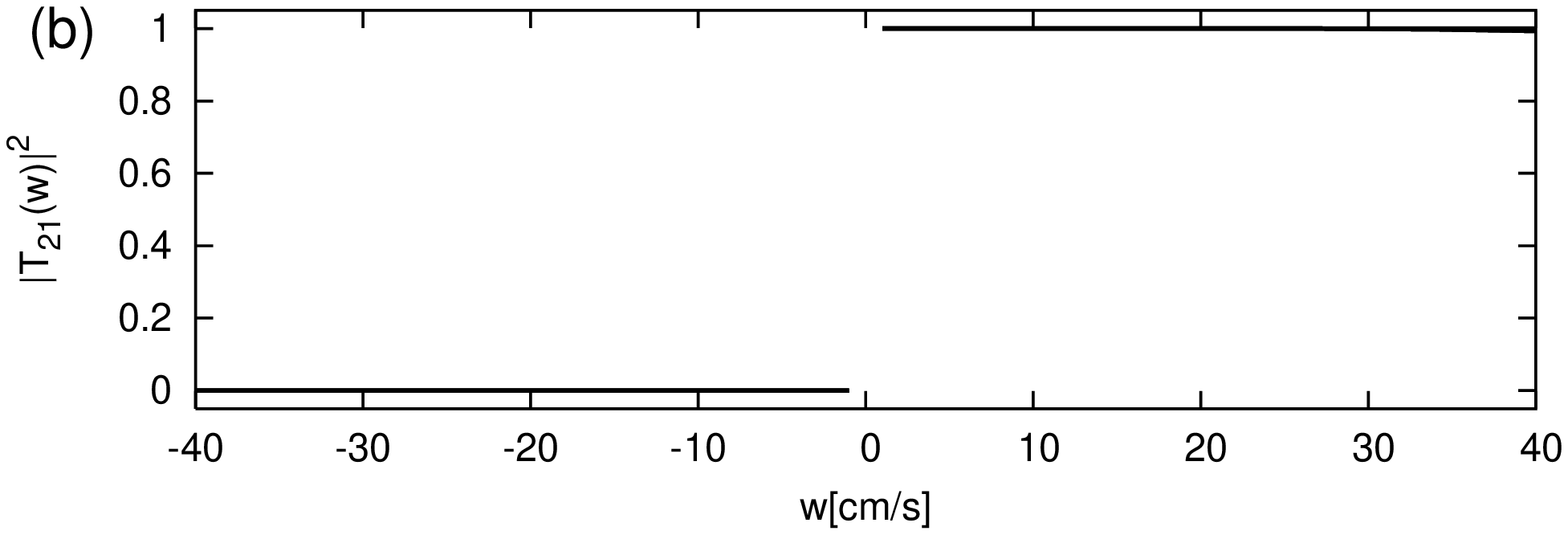}
\end{center}
\caption{\label{fig2}(a) Reflection probability $\fabsq{R_{11}(w)}$ and
(b) transmission probability $\fabsq{T_{21} (w)}$; recall that in this 
notation 
$w<0$ corresponds to incidence from the right, and $w>0$ to 
incidence from the left; $d = 50 \mum$;
$\hat{\Omega} = 1 \Msi$, $\hat{W}_1 = \hat{W}_2 = 100 \Msi$;
$\hat{\Omega} = 0.2 \Msi$, $\hat{W}_1 = 20 \Msi$, $\hat{W}_2 = 100 \Msi$;
$\hat{\Omega} = 0.2 \Msi$, $\hat{W}_1 = 100 \Msi$, $\hat{W}_2 = 20 \Msi$
(coinciding solid lines).}
\end{figure}
%
The behavior of the two-level atom diode is examined by solving numerically 
the
stationary Schr\"odinger equation,
\begin{eqnarray}
E_v \bm{\Psi} (x) = \bm{H} \bm{\Psi} (x),
\label{stat}
\end{eqnarray}
with the Hamiltonian given by Eq. (\ref{ham2}) and $E_v = \frac{m}{2}v^2$.
The results, obtained by the ``invariant imbedding method''
\cite{singer.1982,band.1994},
are shown in Fig. \ref{fig2} for different parameters.
In the plotted velocity range, the ``diodic''
behaviour holds, i.e. Eqs. (\ref{di1}) and (\ref{di2}) are fulfilled.  
(The transmission and reflection
probabilities for incidence in the ground state, 
$|R_{21}^{l/r}|^2$ and $|T_{11}^{l/r}|^2$, which are not shown
in the Figure are zero.) 
The device may be asymmetric,
i.e. even with $\hat{W}_1 \neq \hat{W}_2$
there can be a ``diodic'' behaviour, see some 
examples in Fig. \ref{fig2}.

Note in passing that the device
works as a diode for incidence in the excited state too, 
but in the opposite direction, namely, 
$|T^r_{12}(v)|^2 \approx |R^l_{22}(v)|^2\approx 1$, 
whereas all other probabilities for incidence in the excited state
are approximately zero.

\begin{figure}[t]
\begin{center}
\includegraphics[angle=-90,width=\linewidth]{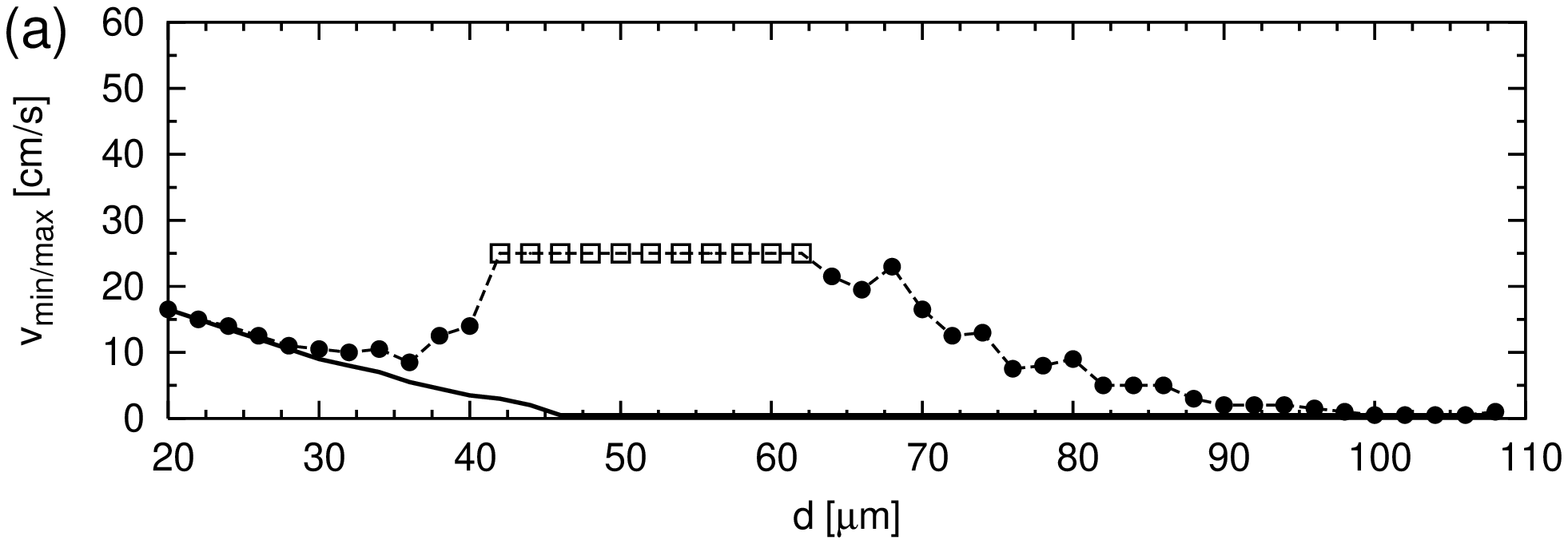}

\includegraphics[angle=-90,width=\linewidth]{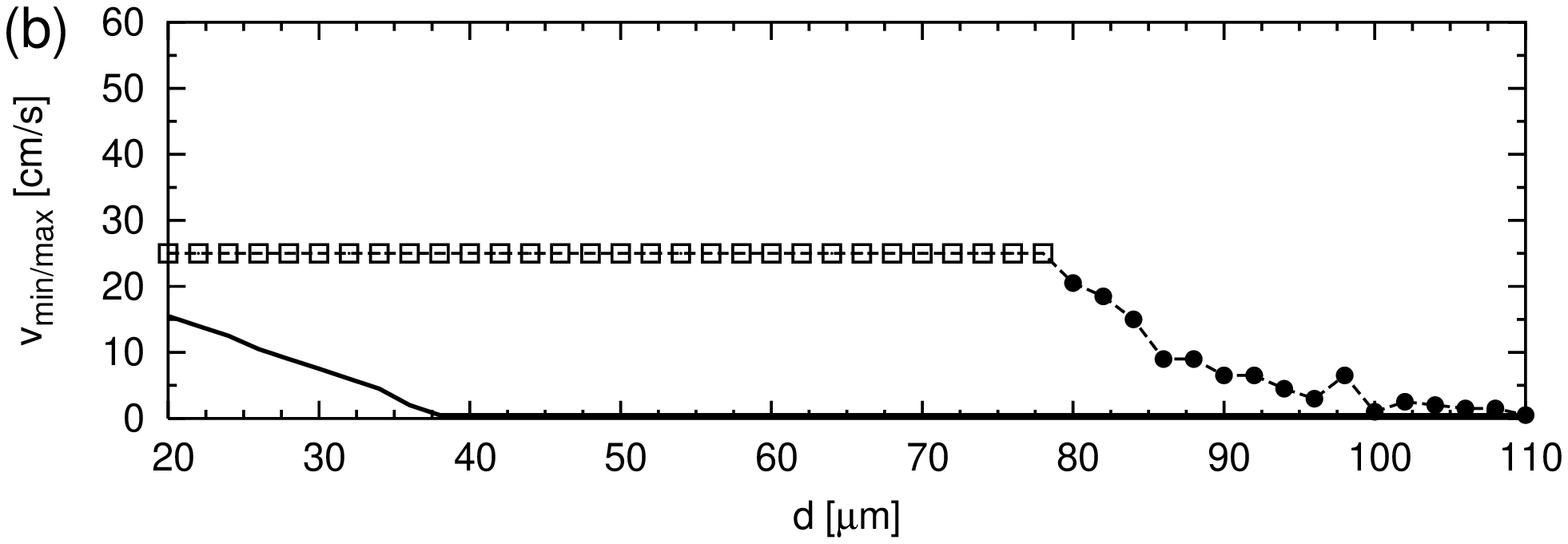}

\includegraphics[angle=-90,width=\linewidth]{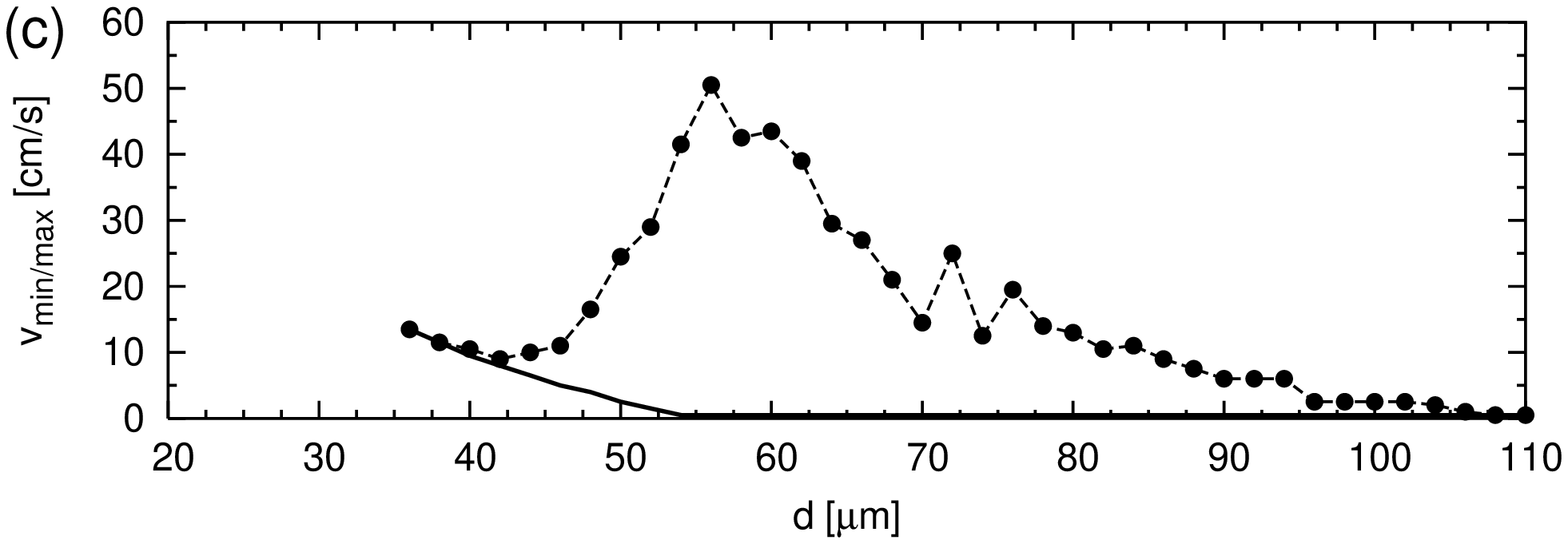}

\includegraphics[angle=-90,width=\linewidth]{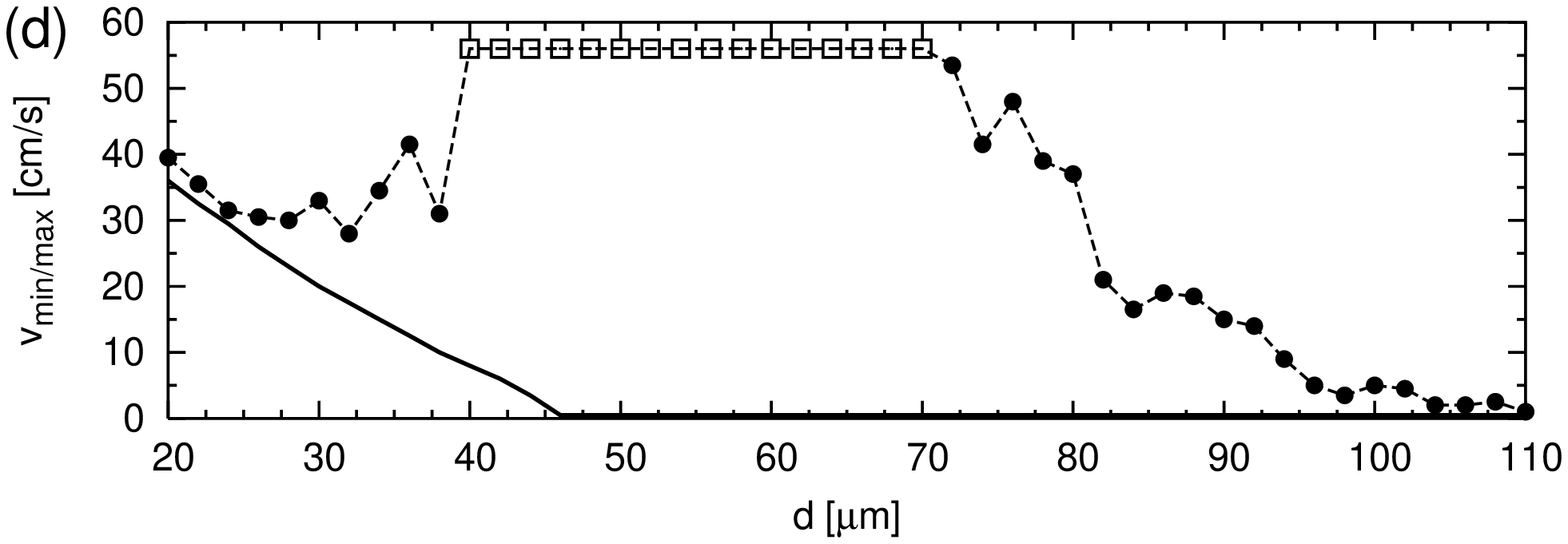}
\end{center}
\caption{\label{fig3} Limit $v_{min}$ (solid lines) and
$v_{max}$ (symbols connected with dashed lines) for ``diodic'' behaviour,
$\epsilon = 0.01$; the circles (boxes) correspond to
breakdown due to transmission (reflection);
(a) $\hat{\Omega} = 0.2 \Msi$, $\hat{W}_1 = \hat{W}_2 = 20 \Msi$;
(b) $\hat{\Omega} = 1 \Msi$, $\hat{W}_1 = \hat{W}_2 = 20 \Msi$;
(c) $\hat{\Omega} = 0.2 \Msi$, $\hat{W}_1 = \hat{W}_2 = 100 \Msi$;
(d) $\hat{\Omega} = 1 \Msi$, $\hat{W}_1 = \hat{W}_2 = 100 \Msi$.}
\end{figure}

Now let us examine the stability of the diode with respect to 
changes in the separation between laser field centers $d$.
We define $v_{max}$ and $v_{min}$ as the upper and lower limits where  
diodic behaviour holds,  
by imposing that all scattering probabilities from the ground state be small 
except the ones in Eq. (\ref{di1}) (i.e., the transmission probability 
from $1$ to $2$ for left  
incidence and the reflection probability from $1$ to $1$ for right incidence).
More precisely, 
$v_{max/min}$ are chosen as the limiting values such that  
$\sum_{\beta=1}^2 (|R_{\beta 1}^l|^2+|T^r_{\beta 1}|^2)
+(|R^r_{21}|^2+|T^l_{11}|^2)+(1-|T_{21}^l|^2)+
(1-|R^r_{11}|^2)<\epsilon$ for all $v_{min}< v <v_{max}$. 
In Fig. \ref{fig3}, $v_{max/min}$ are 
plotted versus the distance between the laser centers, $d$,
for different combinations of $\hat{\Omega}$, $\hat{W}_1$,
and $\hat{W}_2$. For the intensities
considered, $v_{max}$ is in the ultra-cold regime below $1$ m/s. 
In the $v_{max}$ surface, unfilled boxes indicate reflection failure 
for right incidence and filled circles indicate 
transmission failure for left incidence.
In the $v_{min}$ surface, the failure is always due to
transmission failure for left incidence.
We see that the valid $d$ range for ``diodic'' behaviour
can be increased by increasing the Rabi frequency
$\hat{\Omega}$, compare e.g. (a) and (b), 
or (c) and (d). 
Moreover, higher mirror intensities increase $v_{max}$ at the plateau 
but also make it narrower, compare e.g. (b) and (d). 
This narrowing can be simply compensated by increasing $\hat{\Omega}$ too, 
compare  e.g. (a) and (d). 

%

%
\begin{figure}[t]
\begin{center}
\includegraphics[angle=-90,width=\linewidth]{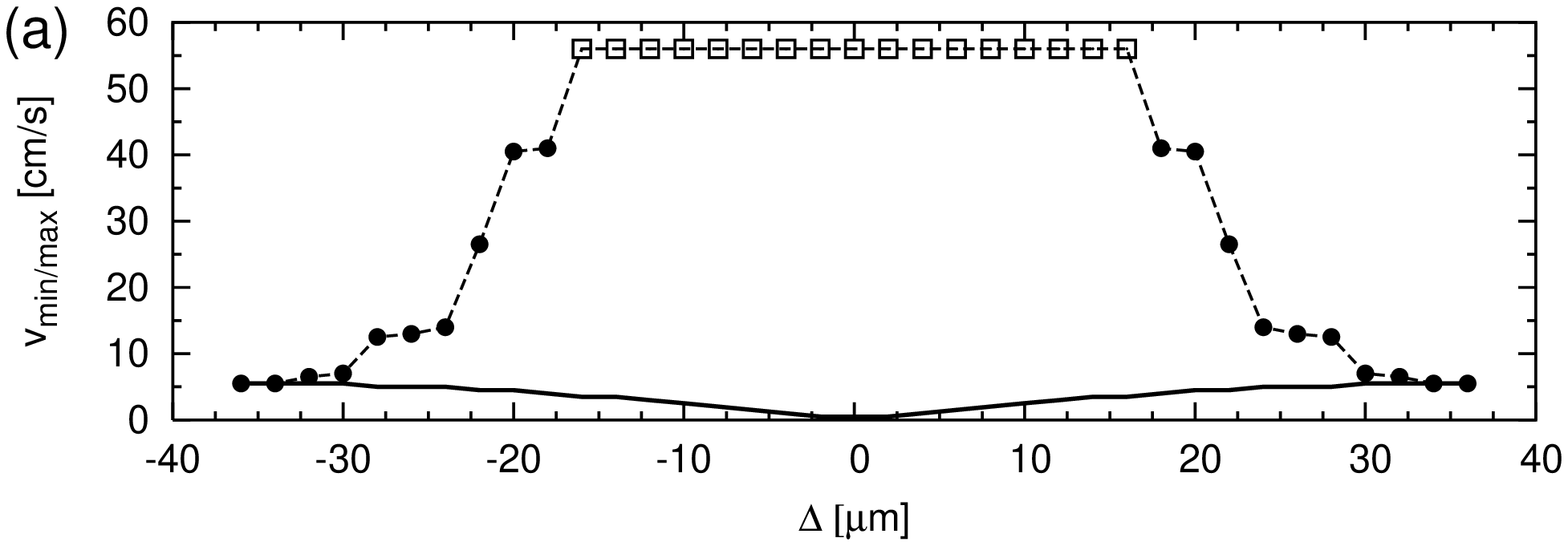}

\includegraphics[angle=-90,width=\linewidth]{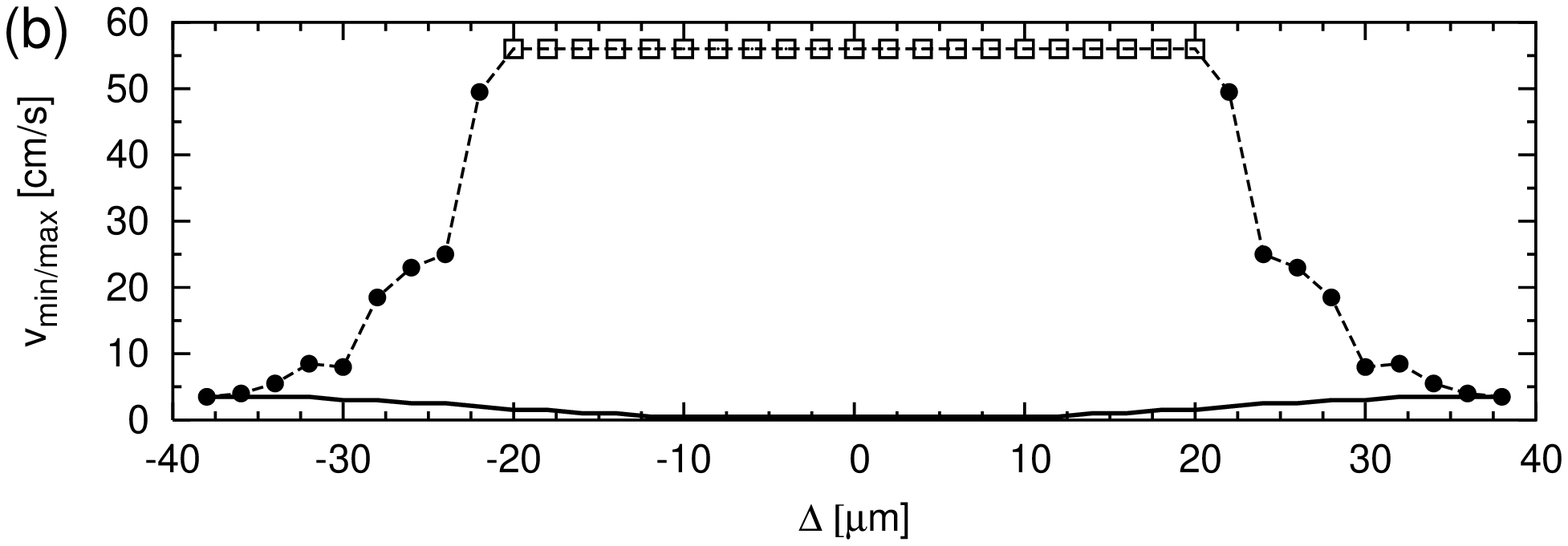}

\includegraphics[angle=-90,width=\linewidth]{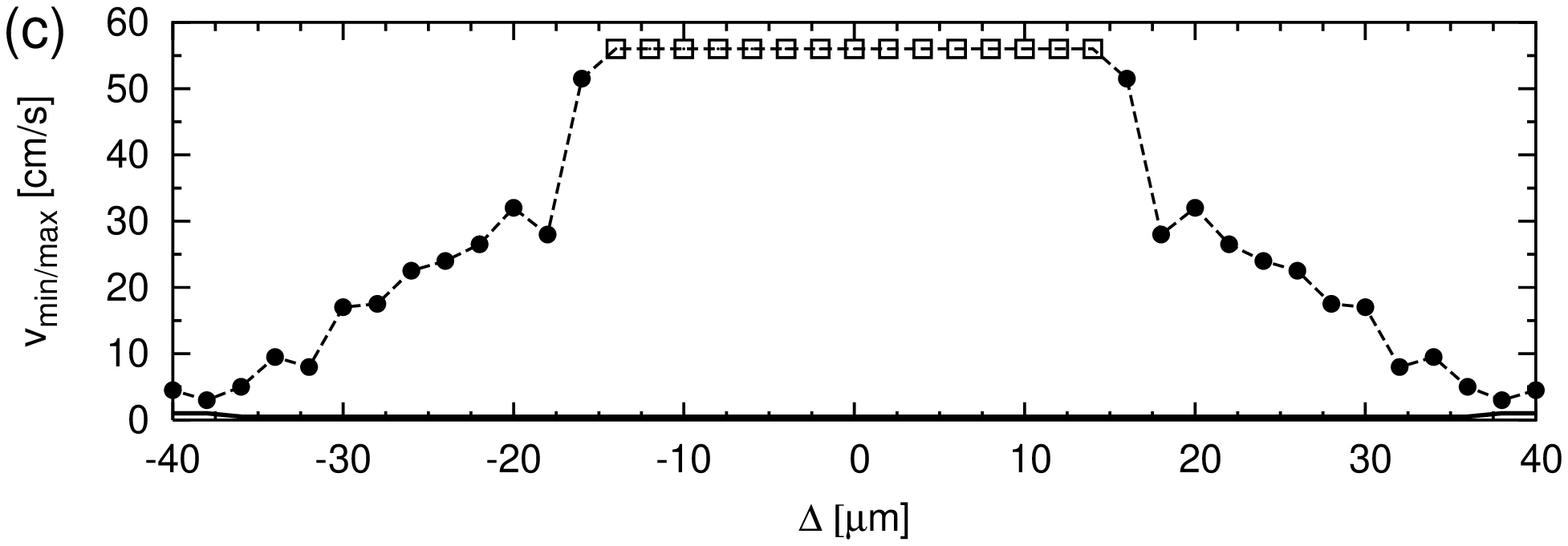}

\includegraphics[angle=-90,width=\linewidth]{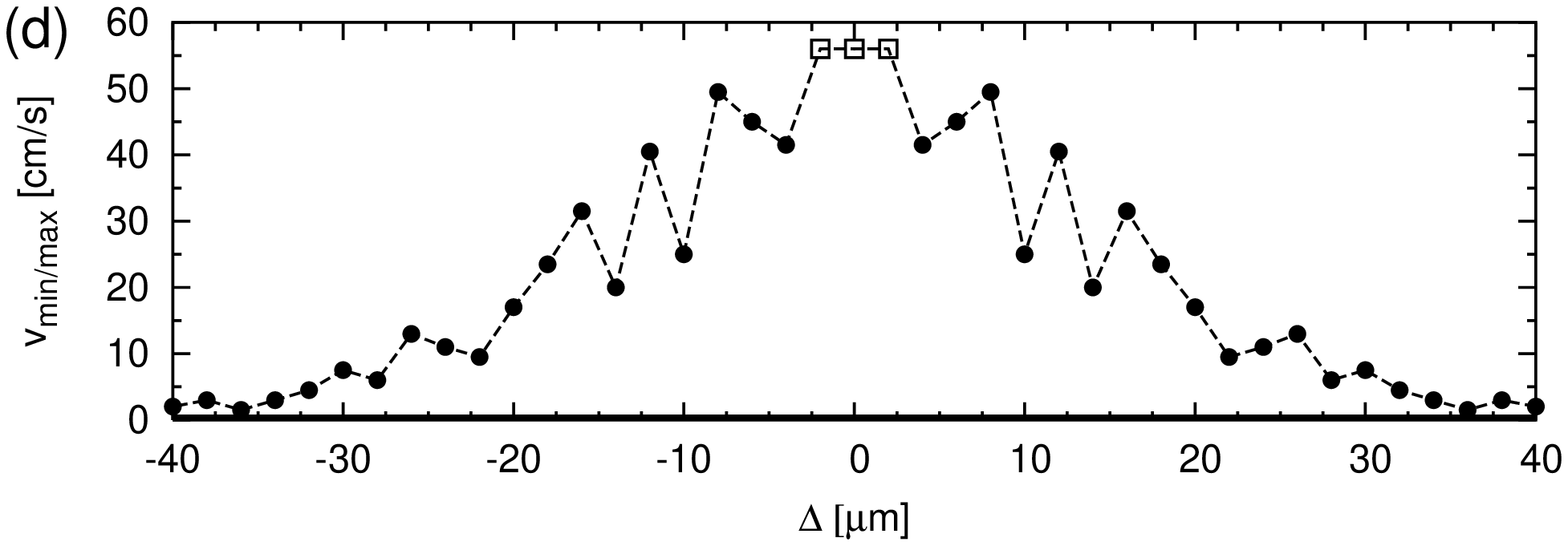}
\end{center}
\caption{\label{figx} Limit $v_{min}$ (solid lines) and
$v_{max}$ (symbols connected with dashed lines) for ``diodic'' behaviour
versus the shift $\Delta$ for different $d$,
$\epsilon = 0.01$; the circles (boxes) correspond to
breakdown first for transmission (reflection);
$\hat{\Omega} = 1 \Msi$, $\hat{W}_1 = \hat{W}_2 = 100 \Msi$;
(a) $d=46\mum$,
(b) $d=50\mum$,
(c) $d=60\mum$,
(d) $d=70\mum$.}
\end{figure}

Finally, we have also examined the stability with respect to 
a shift $\Delta$
of the central position of the pumping laser, 
see Fig. \ref{figx}.
It turns out that there is a range, which depends on $d$, 
where the limits $v_{min}$ and $v_{max}$
practically do not change. 
\section{Variants of the atom diode\label{s3}}
\begin{figure}[t]
\begin{center}
\includegraphics[angle=-90,width=0.95\linewidth]{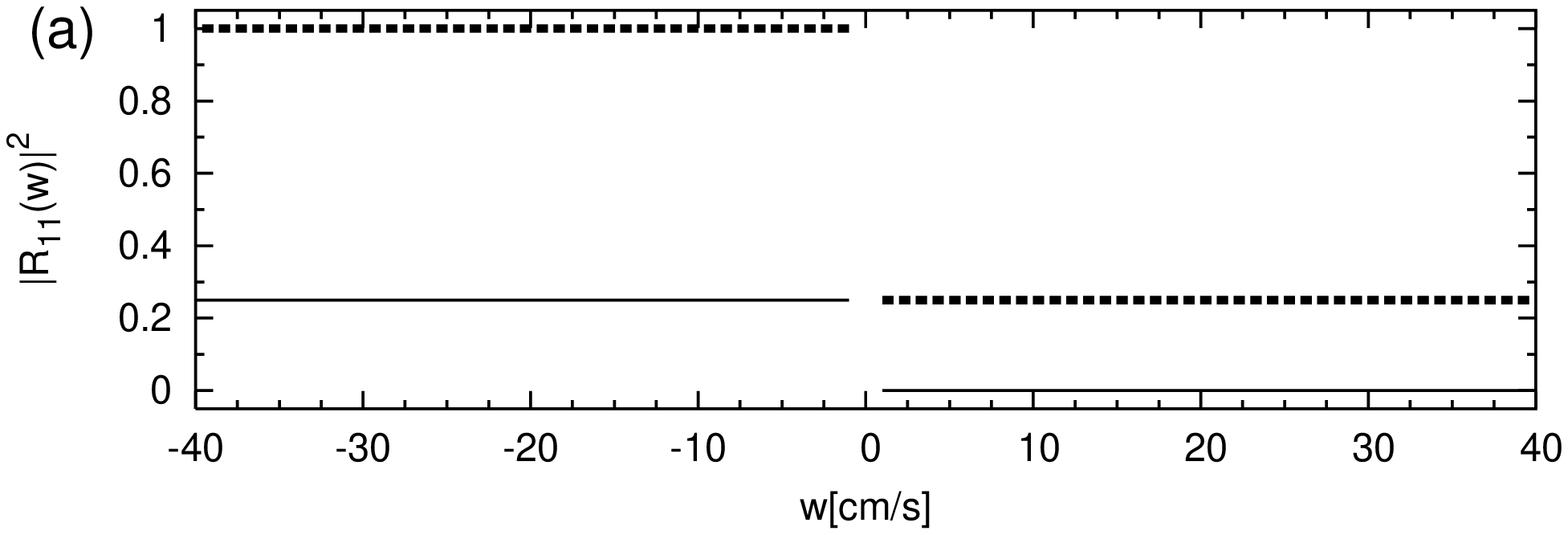}

\includegraphics[angle=-90,width=0.95\linewidth]{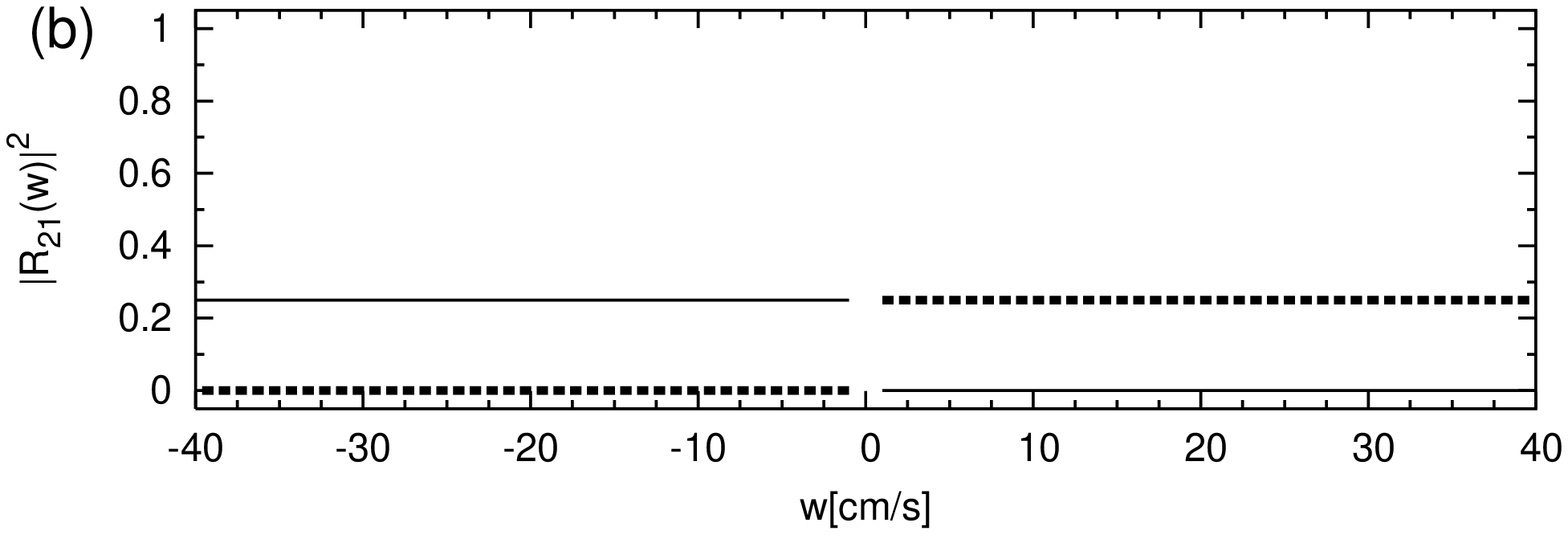}

\includegraphics[angle=-90,width=0.95\linewidth]{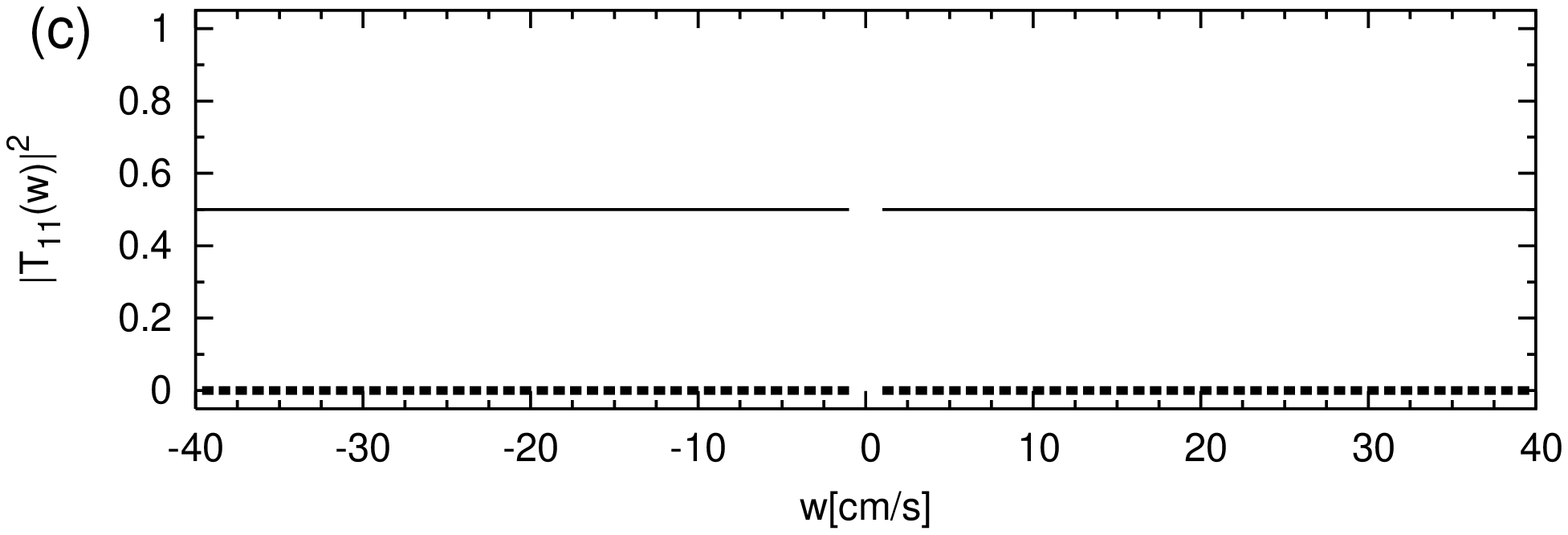}

\includegraphics[angle=-90,width=0.95\linewidth]{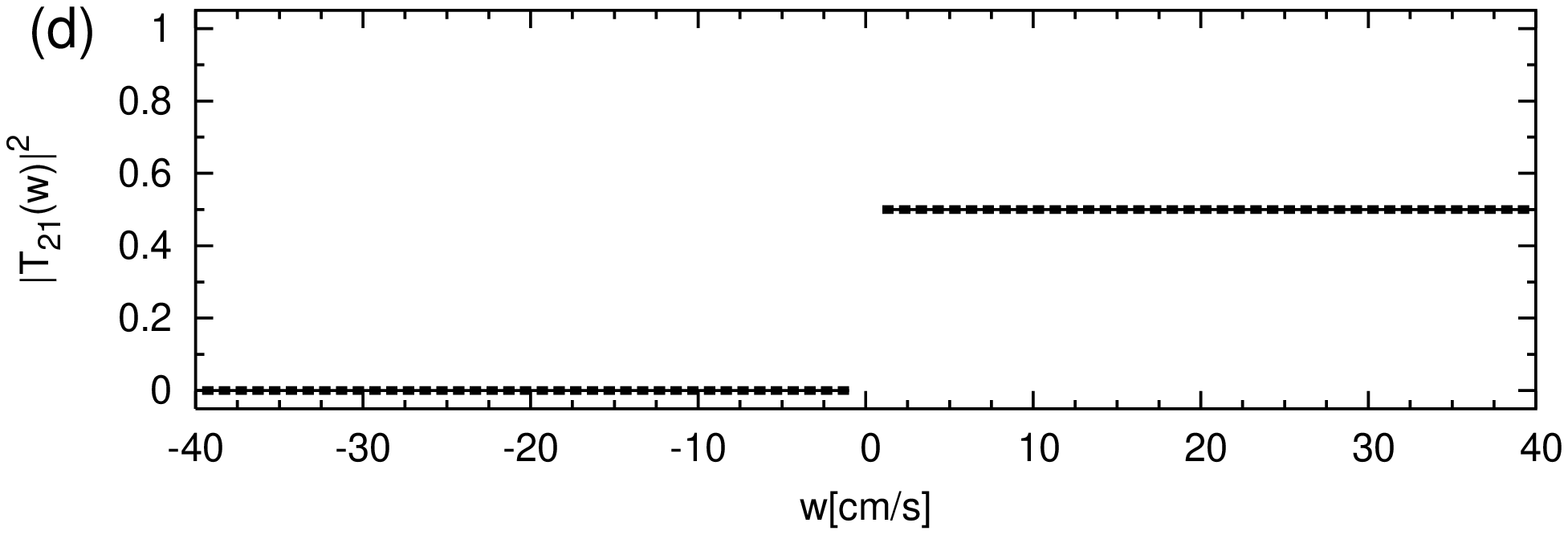}
\end{center}
\caption{\label{fig5}(a) Reflection probability $\fabsq{R_{11}(w)}$,
(b) reflection probability $|{R_{21}(w)}|^2$, 
(c) transmission probability $\fabsq{T_{11} (w)}$;
(d) transmission probability $\fabsq{T_{21} (w)}$;
$d = 50 \mum$, $\hat{\Omega} = 1 \Msi$;
$\hat{W}_1 = 100 \Msi$, $\hat{W}_2 = 0$ (dashed lines);
$\hat{W}_1 = 0$, $\hat{W}_2 = 100 \Msi$ (solid lines).}
\end{figure}

Is the mirror potential $W_2$ really necessary?
If we want ground state atoms to pass from left to right
but not from right to left,  
it is not intuitively obvious why we should add a reflection 
potential for the excited state on the left of the pumping potential 
$\Omega$, see again Fig. \ref{fig1}. 
In other words, it could appear that the pumping potential 
and a reflecting potential $W_1$ on its right 
would be enough to make  a perfect diode.
This simpler two-laser scheme, however, only works 
partially.   
In Fig. \ref{fig5} the scattering probabilities for the case
$\hat{W}_1 >0$, $\hat{W}_2 = 0$ are represented.
While there is still full reflection if the
atom comes from the right, the transmission probability 
is only
$1/2$ when the atom comes from the left; accordingly there is a $1/2$ reflection 
probability from the left,   
which is equally distributed between the ground
and excited state channels.
This is in contrast to the $\hat{W}_1>0$, $\hat{W}_2 > 0$ case of Fig \ref{fig2}.
We may thus conclude 
that the counterintuitive state-selective mirror
$W_2$ is really 
important to attain  
a perfect diode.

\begin{figure}[t]
\begin{center}
\includegraphics[angle=-90,width=0.95\linewidth]{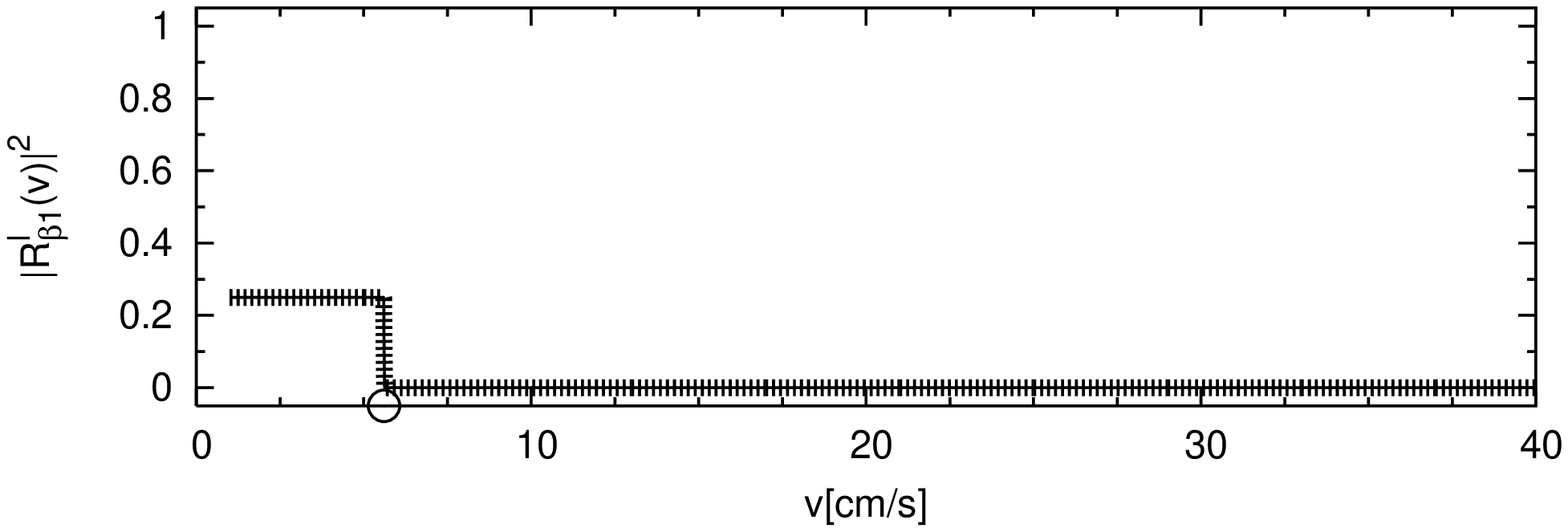}

\includegraphics[angle=-90,width=0.95\linewidth]{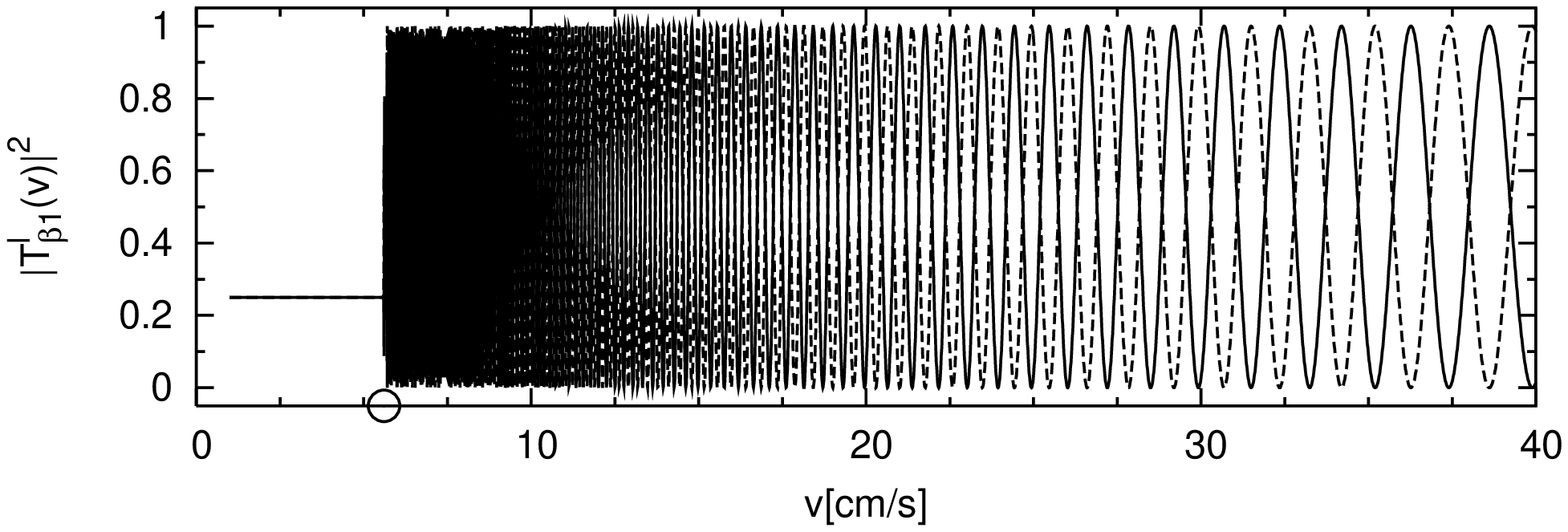}
\end{center}
\caption{\label{fig6}Reflection and transmission probability
for incidence from the left, $d = 50 \mum$, $\hat{\Omega} = 1 \Msi$,
$\hat{W}_1 = \hat{W}_2 = 0$;
the circles indicate $v_{\lambda,max}$ while in this
case $v_{\lambda,min}=0$ (see Eqs. (\ref{deflm}) and (\ref{deflp}));
(a) $\fabsq{R^l_{11}(v)}$ (thick dotted line),
$\fabsq{R^l_{21}(v)}$ (solid line);
(b) $\fabsq{T^l_{11} (v)}$ (dashed line),
$\fabsq{T^l_{21} (v)}$ (solid line).}
\end{figure}

In Fig. \ref{fig5} the case $\hat{W}_1=0$, $\hat{W}_2>0$ is also plotted.
For incidence from the right in the ground state 
there is no full reflection so this case
is not useful as a diode.
But for incidence from the left there is equal transmission in
ground and excited states so that this device might be useful
to build an interferometer.
A very remarkable and useful property in this case, and in
fact in all cases depicted in 
Figs. \ref{fig2} and \ref{fig5}, is the constant value of the 
transmission and reflection
probabilities in a broad velocity range. This is calling for an explanation. 
Moreover, why do they take the values $1$, $1/2$, or $1/4$? 
None of these facts is very intuitive, neither within the 
representations and concepts we have put forward so far, nor 
according to the following arguments: 
Let us consider 
again the simple 
two-laser configuration with $\hat{W}_2=0$ and $\hat{W}_1>0$. 
From a classical perspective, 
the atom incident from the left finds first the pumping laser and then 
the state-selective mirror potential for the ground state. According 
to this ``sequential'' model,  
one would expect an important effect of the velocity in the pumping 
efficiency. A different velocity implies a different traversal time and thus 
a different final phase for the Rabi oscillation which should  
lead  to  
a smooth, continuum variation of the final atomic state 
with the velocity. In particular, 
the probability of the excited 
state after the pumping would oscillate with the velocity and therefore the 
final transmission after the right mirror should oscillate too,  
if the sequential model picture were valid. 
Indeed, these 
oscillations are  clearly seen in Fig. \ref{fig6} 
when $\hat{W}_1=\hat{W}_2=0$
(above a low velocity threshold 
in which the Rabi oscillations are suppressed and all channels are equally 
populated, for a related effect see   
\cite{ROS}, see also the explanation of this low
velocity regime in section \ref{s4}).
Clearly, however, the oscillations  
are absent when $\hat{W}_1>0$, so the sequential, classical-like 
picture cannot be right. In summary, the mirror potentials 
added to the pumping laser 
imply a noteworthy stabilization of
the probabilities and velocity independence. 
The failure of the sequential scattering picture must be due to 
some sort of quantum interference phenomenon. Interference effects 
are well known in scattering off  
composite potentials, but in comparison with, e.g., resonance peaks 
in a double barrier, the present results are of a different nature. 
There is indeed a clean explanation to all the mysteries we have 
dropped along the way as the reader
will find out in the next section.

%
%

%
%
%
%
%
%
%
%
\section{Adiabatic interpretation of the diode and its variants\label{s4}}
Depending on the mirror potentials included 
in the device, let us label the four possible cases discussed in the 
previous section as follows:  
case ``0'': $\hat{W}_1=\hat{W}_2=0$; 
case ``1'': $\hat{W}_1>0$, $\hat{W}_2=0$; 
case ``2'': $\hat{W}_1=0$, $\hat{W}_2>0$; 
case ``12'': $\hat{W}_1>0$, $\hat{W}_2>0$.
We diagonalize now the potential matrix $\bm{M}(x)$
\begin{eqnarray*}
\bm{U}(x)\bm{M}(x)\bm{U}^+ (x) = \left(\begin{array}{cc}
\lambda_-(x) & 0 \\ 0 & \lambda_+(x)
\end{array}\right). 
\end{eqnarray*}
The orthogonal matrix $\bm{U}(x)$ is given by
\begin{eqnarray*}
\bm{U}(x) = \left(\begin{array}{cc}
\frac{W_-(x) - \mu(x)}{\sqrt{4\Omega^2(x)+[W_-(x) - \mu(x)]^2}} &
\frac{W_-(x) + \mu(x)}{\sqrt{4\Omega^2(x)+[W_-(x) + \mu(x)]^2}}\\
\frac{2\Omega(x)}{\sqrt{4\Omega^2(x)+[W_-(x) - \mu(x)]^2}} &
\frac{2\Omega(x)}{\sqrt{4\Omega^2(x)+[W_-(x) + \mu(x)]^2}}
\end{array}\right)
\end{eqnarray*}
where  
\beqa
W_- &=& W_1 - W_2, 
\nonumber\\
\mu&=&\sqrt{4\Omega^2(x)+W_-^2(x)}, 
\nonumber
\eeqa
and the eigenvalues of $\bm{M}(x)$ are 
$$
\lambda_{\mp} (x) = \frac{\hbar}{4}\left[W_1(x)+W_2(x) \mp \mu(x)\right] 
$$
with corresponding (normalized) eigenvectors $|\lambda_\mp(x)\ra$. 
The asymptotic form of $\bm{U}$ varies for the different cases
distinguished with a superscript, $U^{(j)}$,
$j=0,1,2,12$.  
For $x\to-\infty$, the same $\bm{U}$ is found for cases $0$ and $1$,
in which the left edge corresponds to the pumping potential. 
Similarly, the cases $2$ and $12$ share the same left edge potential 
$W_2$ and thus a common form of $\bm{U}$,    
\begin{eqnarray*}
\bm{U}^{(0,1)}(-\infty)\!=\!\frac{1}{\sqrt{2}}\left(\begin{array}{cc}
-1 & 1\\
1 & 1
\end{array}\right),\quad\!
\!\!\bm{U}^{(2,12)}(-\infty)\!=\!\left(\begin{array}{cc}
-1 & 0\\
0 & 1
\end{array}\right)\!.
\end{eqnarray*}
The corresponding analysis for $x\to \infty$ gives the asymptotic 
forms   
\begin{eqnarray*}
\bm{U}^{(0,2)}(\infty) = \frac{1}{\sqrt{2}}\left(\begin{array}{cc}
-1 & 1\\
1 & 1
\end{array}\right),\quad
\bm{U}^{(1,12)}(\infty) = \left(\begin{array}{cc}
0 & 1\\
1 & 0
\end{array}\right).
\end{eqnarray*}
%
%
\begin{figure}[t]
\begin{center}
\includegraphics[angle=-90,width=\linewidth]{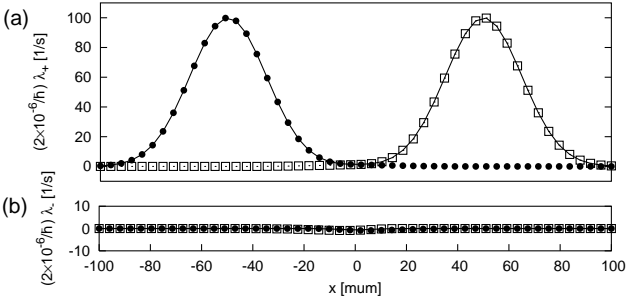}
\end{center}
\caption{\label{fig7} Eigenvalues (a) $\lambda_+$ and (b) $\lambda_-$;
$d=50\mum$, $\hat{\Omega} = 1 \Msi$;
$\hat{W}_1 = \hat{W}_2 = 100 \Msi$ (solid lines);
$\hat{W}_1 = 100 \Msi$, $\hat{W}_2 = 0$ (boxes);
$\hat{W}_1 = 0$, $\hat{W}_2 = 100 \Msi$ (circles).}
\end{figure}
%
The eigenvalues $\lambda_\pm (x)$ for the same parameters
of Fig. \ref{fig5} are plotted in Fig. \ref{fig7}.
We see that $\lambda_+ (x) > 0$
has at least one high barrier whereas $\lambda_-(x) \approx 0$.

If $\bm{\Psi}$ is a two-component 
solution of the stationary
Schr\"odinger equation, Eq.  (\ref{stat}),  
we define now the vector 
$$
\bm{\Phi}(x)= {\phi_-(x) \choose \phi_+(x)} := 
\bm{U}(x)\bm{\Psi}(x)
$$
in a potential-adapted, ``adiabatic representation''.
Note that if no approximation is made, $\bm{\Phi}$ and $\bm{\Psi}$ are 
both exact 
and contain the same information expressed in different bases.  
Starting from Eq. (\ref{stat}),
using $\bm{\Psi}=\bm{U}^+\bm{\Phi}$,
and multiplying 
from the left by $\bm{U}$, we  arrive
at the following equation for $\bm{\Phi}(x)$
\begin{eqnarray*}
E_v \bm{\Phi}(x) &=&
-\frac{\hbar^2}{2m} \frac{\partial^2}{\partial x^2} \bm{\Phi}(x)
+ \left(\begin{array}{cc}
\!\lambda_-(x) & 0 \\ 0 & \lambda_+(x)
\end{array}\!\right) \bm{\Phi} (x)\\
& & + \bm{Q} \bm{\Phi}(x),
\end{eqnarray*}
where
\begin{eqnarray}
\bm{Q} &=& -\frac{\hbar^2}{2m} \left(\bm{U}(x)
\frac{\partial^2 \bm{U}^+}{\partial x^2}(x)
+ 2 \bm{U}(x)\frac{\partial \bm{U}^+}{\partial x}(x)
\frac{\partial}{\partial x}\right)
\nonumber
\\
& = & \left(\begin{array}{cc}
m\,B^2(x)/2 & -A(x) + i B(x) \hat{p}_x
\\
A(x) - i B(x) \hat{p}_x & m\,B^2(x)/2\end{array}\right)
\label{q}
\end{eqnarray}
is the coupling term in the adiabatic basis,  
and $A(x)$, $B(x)$ are real functions, 
\begin{eqnarray*}
A(x)&=& \frac{1}{32 \mu^4(x)\Delta x^4 m} \Big\{\\
& & \fexp{-\frac{(x+d)^2}{\Delta x^2}} d^2 \hbar^6 \Omega(x) W_-(x)\\
& & + \fexp{\frac{(x+d)^2}{\Delta x^2}} \times\\
& & \left[-4\Omega^2(x) + W_1^2(x) + W_2^2(x)
+ 6 W_1(x)W_2(x)\right]\Big\},\\
B(x)&=& \frac{d \hbar^3}{4 \mu^2(x) \Delta x^2 m} \Omega(x)
\left[W_1(x) + W_2(x)\right]. 
\end{eqnarray*}
Let us consider incidence from the left
and assume first that the coupling $\bm{Q}$ can be neglected so that there are 
two independent adiabatic modes ($\pm$)  
in which the internal state of the atom 
adapts to the position-dependent eigenstates $|\lambda_\pm\ra$ 
of the laser potential $\bm{M}$,
whereas the 
atom center-of-mass motion is affected in each mode by the effective 
adiabatic potentials $\lambda_\pm(x)$.   

Because $\lambda_- \approx 0$, an approximate
solution for $\phi_- (x)$ is a full
transmitted wave and because $\lambda_+$ consists of at least one ``high''
barrier -at any rate the present argument is only applicable for energies below the 
barrier top-,
an approximate solution for $\phi_- (x)$ is a wave which is fully reflected by
a wall.
So we can write for $x\ll 0$, 
\begin{eqnarray*}
\bm{\Phi} (x) \approx \bm{\Phi}_{-\infty} (x) := 
\left(\begin{array}{c}c_-\\c_+\end{array}\right) e^{ikx}
+ \left(\begin{array}{c}0\\-c_+\end{array}\right) e^{-ikx},
\end{eqnarray*}
and for $x\gg 0$, 
\begin{eqnarray*}
\bm{\Phi} (x) \approx \bm{\Phi}_{\infty} (x) :=
\left(\begin{array}{c}c_-\\0\end{array}\right) e^{ikx}. 
\end{eqnarray*}
In order to determine the amplitudes $c_\pm$ we have to compare with 
the asymptotic form of  
the scattering solution for left incidence,   
\begin{eqnarray*}
{\bm{\Psi}}(x) \approx {\bm{\Psi}}_{-\infty} (x) :=
 \left(\begin{array}{c} 1\\0 \end{array}\right) e^{i k x} 
+ \left(\begin{array}{c} R_{11}^l\\R_{21}^l \end{array}\right) e^{-i k x} 
\end{eqnarray*}
if $x\ll 0$ and
\begin{eqnarray*}
\bm{\Psi}(x) \approx \bm{\Psi}_{\infty} (x) := e^{i k x}
\left(\begin{array}{c} T_{11}^l\\T_{21}^l \end{array}\right)
\end{eqnarray*}
if $x\gg 0$.

The transmission and reflection coefficients can now be approximately calculated
for each case from the boundary conditions
$\bm{\Phi}_{-\infty}(x) = \bm{U}(-\infty) \bm{\Psi}_{-\infty} (x)$ and
$\bm{\Phi}_{\infty}(x) = \bm{U}(\infty) \bm{\Psi}_{\infty} (x)$.

\begin{table}[tbp]
\label{tab}
\caption{Reflection and transmission probability for the different
variations of the atom diode}
(a) incidence from the right:
\begin{eqnarray*}
\begin{array}{cc|c|c|c|c|c|c|}
&\mbox{case} & c_-^{r} & c_+^{r} & R_{11}^{r}& R_{21}^{r} &
T_{11}^{r} & T_{21}^{r}\\[0.1cm]
\hline
(0)& \hat{W}_1=\hat{W}_2=0 & -\frac{1}{\sqrt{2}} & \frac{1}{\sqrt{2}} &
-\frac{1}{2} & -\frac{1}{2} & \frac{1}{2} & -\frac{1}{2}\\[0.1cm]
\hline
(1)& \hat{W}_1>0, \hat{W}_2=0 & 0 & 1 &
-1 & 0 & 0 & 0\\[0.1cm]
\hline
(2)& \hat{W}_1=0, \hat{W}_2>0 & -\frac{1}{\sqrt{2}} & \frac{1}{\sqrt{2}} &
-\frac{1}{2} & -\frac{1}{2} & \frac{1}{\sqrt{2}} & 0\\[0.1cm]
\hline
(12)& \hat{W}_1>0, \hat{W}_2>0 & 0 & 1 &
 -1 & 0 & 0 & 0\\[0.1cm]
\hline
\end{array}
\end{eqnarray*}

(b) incidence from the left:\vspace{-0.2cm}
\begin{eqnarray*}
\begin{array}{cc|c|c|c|c|c|c|}
&\mbox{case} & c_-^{l} & c_+^{l} & R_{11}^{l}& R_{21}^{l} &
T_{11}^{l} & T_{21}^{l}\\[0.1cm]
\hline
(0)& \hat{W}_1=\hat{W}_2=0 & -\frac{1}{\sqrt{2}} & \frac{1}{\sqrt{2}} &
-\frac{1}{2} & -\frac{1}{2} & \frac{1}{2} & -\frac{1}{2}\\[0.1cm]
\hline
(1)& \hat{W}_1>0, \hat{W}_2=0 & -\frac{1}{\sqrt{2}} & \frac{1}{\sqrt{2}} &
-\frac{1}{2} & -\frac{1}{2} & 0 & -\frac{1}{\sqrt{2}}\\[0.1cm]
\hline
(2)& \hat{W}_1=0, \hat{W}_2>0 & -1 & 0 &
0 & 0 & \frac{1}{\sqrt{2}} & -\frac{1}{\sqrt{2}}\\[0.1cm]
\hline
(12)& \hat{W}_1>0, \hat{W}_2>0 & -1 & 0 &
0 & 0 & 0 & -1\\[0.1cm]
\hline
\end{array}
\end{eqnarray*}
\end{table}

The incidence from the right can be treated in a similar way.
All the amplitudes are given in Table I, from which we can find, taking the 
squares, the transmission and reflection probabilities $1$, $1/2$, $1/4$, 
and $0$,  
of Figs. \ref{fig2}, \ref{fig5}, and \ref{fig6}.

%
\begin{figure}[t]
\begin{center}
\includegraphics[angle=-90,width=\linewidth]{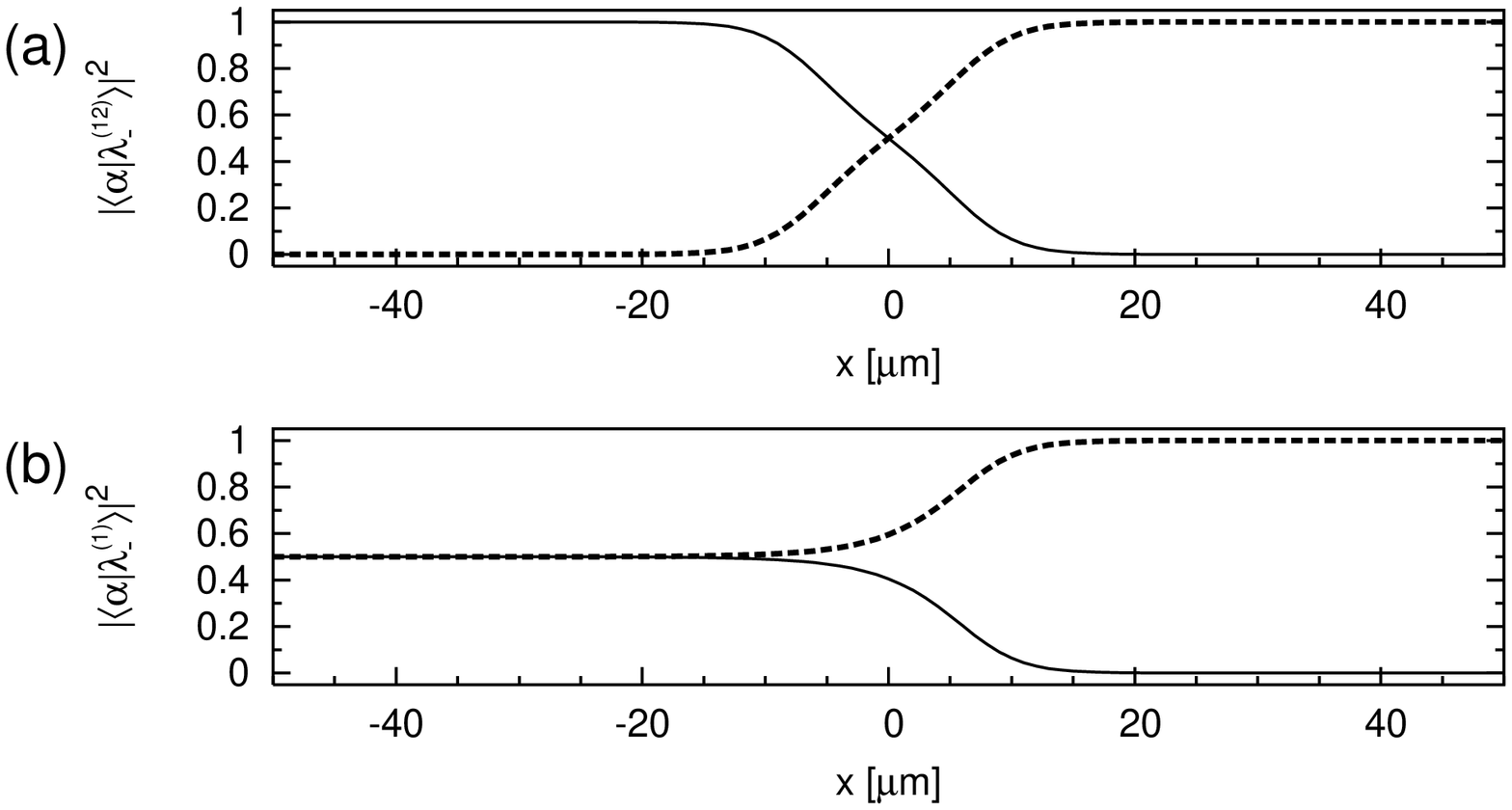}
\end{center}
\caption{\label{fig8} $|\la 1|\lambda_-^{(j)}\ra|^2$ (solid lines)
and $|\la 2|\lambda_-^{(j)}\ra|^2$ (dashed lines) for
$d = 50 \mum$, $\hat{\Omega} = 1 \Msi$, $\hat{W}_1 = 100 \Msi$;
(a) $\hat{W}_2 = 100 \Msi$ (case $j=12$), (b) $\hat{W}_2 = 0$ (case $j=1$).}
\end{figure}
%
These results provide in summary a simple explanation of the 
behaviour of the diode and its variants. In particular, the perfect
diode behavior of case $12$,
occurs because 
the (approximately) ``freely'' moving mode $\phi_-$ transfers
adiabatically the ground state to  
the excited state from left to right.
To visualize this, let
us represent the probabilities to find the ground and excited state 
in the eigenvectors $|\lambda^{(j)}_-(x)\ra$ for the cases 
$j=12,1$.    
They are 
plotted in Fig. \ref{fig8}a
for the case ``12'': the perfect adiabatic 
transfer can be seen clearly.
On the other hand, the mode ``$+$'' (not plotted), which tends  
to the ground state on the right edge of the device, is blocked by a high 
barrier. The stability of this blocking effect with respect 
to incident velocities holds for energies smaller than the $\lambda_+$ barrier
top, more on this below.
In Fig. \ref{fig8}b the ground and excited state probabilities for 
case ``1'' are plotted.
If the mirror potential laser $W_2$
is removed on the left edge of the
device, the ground state is not any more 
an eigenstate of the potential for $x\ll 0$. The adiabatic transfer 
of the mode ``$-$'' occurs instead from $(|2\ra-|1\ra)/2^{1/2}$ on the left 
to $|2\ra$ on the right, whereas the blocked mode ``$+$'' on the left 
corresponds to the linear combination  $(|2\ra+|1\ra)/2^{1/2}$. This results in 
a $1/2$ reflection probability for ground-state incidence from the left. 
A similar analysis would be applicable in the other cases.

%
\begin{figure}[t]
\begin{center}
\includegraphics[angle=-90,width=\linewidth]{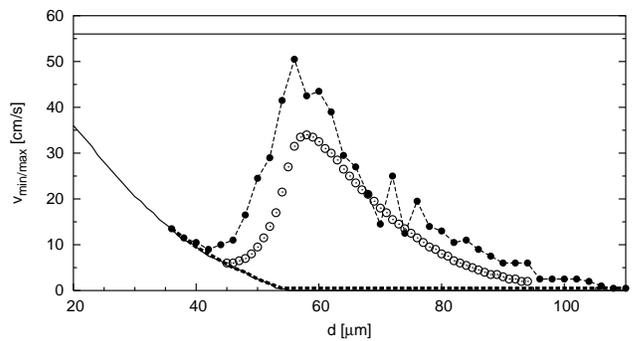}
\end{center}
\caption{\label{fig9} Limits of the ``diodic'' behaviour
$v_{min}$ (thick dashed line) and
$v_{max}$ (filled circles connected with a dashed line, see also Fig. \ref{fig3}),
$\epsilon = 0.01$; limits of condition (\ref{cond1})
$v_{\lambda,min}$ (lower solid line) and $v_{\lambda,max}$ (upper solid line);
limit of the adiabatic approximation $v_{ad,max}$ (unfilled circles),
$\epsilon = 0.01$;
$\hat{W}_1 = \hat{W}_2 = 100 \Msi$, $\hat{\Omega} = 0.2 \Msi$.}
\end{figure}

Of course
all approximations have a range of validity that 
depend on the potential parameters and 
determines the working conditions of the diode. 
Even though these conditions can be easily found numerically from the 
exact results, approximate breakdown criteria 
are helpful to understand the limits of the device  
and different reasons for its failure. 

For the approximation that
$\phi_-$ is a fully transmitted wave and $\phi_+$ a fully
reflected one
a necessary condition is
\begin{eqnarray}
\mbox{max}_x \left[\lambda_-(x)\right] < E_v <
\mbox{max}_x \left[\lambda_+ (x)\right].
\label{cond1}
\end{eqnarray}
This defines the limits
\begin{eqnarray}
v_{\lambda,min} &:=& \sqrt{\frac{2}{m}\mbox{max}_x \left[ \lambda_-(x)\right]},
\label{deflm}\\
v_{\lambda,max} &:=& \sqrt{\frac{2}{m}\mbox{max}_x \left[\lambda_+(x)\right]},
\label{deflp}
\end{eqnarray}
such that Eq. (\ref{cond1}) is fulfilled for all $v$ with
$v_{\lambda,min} < v < v_{\lambda,max}$.
The plateaus of $v_{max}$ seen e.g. in Fig. \ref{fig3} 
for a range of $d$-values are essentially coincident with 
$v_{\lambda,max}$.  
Fig. \ref{fig9} shows the exact limits $v_{min}$ and $v_{max}$
for the ``diodic'' behaviour, as in Fig. \ref{fig3}c,
and also the limits $v_{\lambda,min}$,
$v_{\lambda,max}$ resulting from
the condition of Eq. (\ref{cond1}). We see that the exact limit
$v_{min}$ coincides essentially 
with $v_{\lambda,min}$ so that the lower ``diodic''
velocity boundary can be understood by the breakdown of the condition
that $\phi_-$ is fully transmitted due to a $\lambda_-$ barrier.
This effect is only relevant for small distances $d$
between the lasers.     

Another reason for the breaking down of the diode 
may be that the adiabatic modes are no longer independent, 
i.e. that $\bm{Q}$, see Eq. (\ref{q}), cannot be neglected.
An approximate criterion 
for adiabaticity, more precisely for neglecting the non-diagonal elements of 
$\bm{Q}$, see the Appendix, is  
\begin{eqnarray}
q(v) &:=& \mbox{max}_{x \in I}
\frac{\fabsq{A(x)} + 2m\fabsq{B(x)}[E_v - \lambda_-(x)]}
{\fabsq{\lambda_+(x)-\lambda_-(x)}}\nonumber\\
&\ll& 1
\label{neglectQ}
\end{eqnarray}
with $I = [-d, d]$. A velocity boundary $v_{ad,max}$ defined by 
$q(v) < \epsilon$ for all $v_{\lambda,min} < v < v_{ad,max}$
is shown in Fig. \ref{fig9}. (Note that the condition of Eq. (\ref{neglectQ})
only makes sense if $E_v > \lambda_-(x)$, i.e. $v_{\lambda,min} < v$.)
We see in Fig. \ref{fig9} that the breakdown of the diode at $v_{max}$
for large $d$ is due to a failure of the
adiabatic approximation.

\section{Summary\label{s5}}
Summarizing, we have studied a two-level model for an ``atom diode'', 
a laser device 
in which ground state atoms can pass in one direction, 
conventionally from left to right, but not in the
opposite direction. The proposed scheme includes three lasers:  
two of them are state-selective mirrors, one for the excited state
on the left,
and the other one for  
the ground state on the right,
whereas the third one -located between the two mirrors-
is a pumping laser on resonance with the atomic transition. 
  
We have shown that the ``diodic'' behaviour is very stable
with respect to atom velocity in a given range, and with respect to  
changes in the distances between the centers of the lasers.  
The inclusion of the laser on the left, 
reflecting the excited state, is somewhat counterintuitive, but 
it is essential for a perfect 
diode effect;  
the absence of this laser leads to a $50\%$  drop in efficiency. 
The stability properties as well as the actual mechanism of the diode
is explained with an adiabatic basis and an adiabatic approximation. 
The diodic transmission is due to the adiabatic transfer of
population from left to right, 
from the ground state to the excited state in a free-motion
adiabatic mode, while 
the other mode is blocked by a barrier.

\begin{acknowledgments}
AR acknowledges support by 
the Ministerio de Educaci\'on y Ciencia. 
This work has been supported by Ministerio de Educaci\'on y Ciencia
(BFM2003-01003),
and 
UPV-EHU (00039.310-15968/2004). 
\end{acknowledgments}
%
%
%
%
%
\begin{appendix}
\section{}
To motivate Eq. (\ref{neglectQ}), see also \cite{messiah.book}, 
let us assume
\begin{eqnarray}
E \bm{\Phi}(x) &=& -\frac{\hbar^2}{2m} \frac{\partial^2}{\partial x^2} \bm{\Phi}(x)
+ \left(\begin{array}{cc}
\lambda_- & 0 \\ 0 & \lambda_+
\end{array}\right) \bm{\Phi} (x)\nonumber\\
& + & \epsilon \left(\begin{array}{cc}
0 & -\tilde{A}+i \tilde{B} \hat{p}_x\\
\tilde{A} - i \tilde{B} \hat{p}_x & 0
\end{array}\right) \bm{\Phi}(x)
\label{ap}
\end{eqnarray}
where $\lambda_\pm$, $\tilde{A}$ and $\tilde{B}$
are real and independent of $x$.
We assume that $E > \lambda_-$ and
that $\epsilon$ is small such that we can treat $\bm{\Phi}$
perturbatively,  
\begin{eqnarray*}
\bm{\Phi}(x) \approx
\left(\begin{array}{c}\phi_{0,-} (x) \\ \phi_{0,+} (x)\end{array}\right) +
\epsilon \left(\begin{array}{c}\phi_{1,-} (x) \\ \phi_{1,+} (x)\end{array}\right)
\end{eqnarray*}
with
\begin{eqnarray*}
\phi_{0,-} (x) &=& \fexp{\frac{i}{\hbar} \sqrt{2m(E-\lambda_-)} x}\\
\phi_{0,+} (x) &=& 0
\end{eqnarray*}
Then it follows from (\ref{ap}) for the first-order correction
\begin{eqnarray*}
\phi_{1,-} & = & 0\\
\phi_{1,+} & = & \left[E- \lambda_+ -\hat{p}^2_x/(2m)\right]^{-1}
(\tilde{A} - i \tilde{B} \hat{p}_x)\; \phi_{0,-}\\
& = & \frac{\tilde{A} - i \tilde{B}
\sqrt{2m(E-\lambda_-)}}{\lambda_- - \lambda_+}\; \phi_{0,-}
\end{eqnarray*}
because $\hat{p}_x\,\phi_{0,-} = \sqrt{2m(E-\lambda_-)}\,\phi_{0,-}$.
If we want to neglect $\phi_+ = 0 + \epsilon\,\phi_{1,+}$ we get the condition
\begin{eqnarray*}
\epsilon^2 \frac{\fabsq{\tilde{A}} + \fabsq{\tilde{B}} 2m (E-\lambda_-)}
{\fabsq{\lambda_- - \lambda_+}} \ll 1.
\end{eqnarray*}
If $\lambda_\pm$, $\tilde{A}$ and $\tilde{B}$
depend on $x$, we may use the condition
\begin{eqnarray*}
\mbox{max}_{x\in I} \frac{\fabsq{\epsilon \tilde{A}(x)} +
\fabsq{\epsilon\tilde{B}(x)} 2m [E-\lambda_-(x)]}
{\fabsq{\lambda_-(x) - \lambda_+(x)}} \ll 1
\end{eqnarray*}
where $I$ is chosen in such a way that the assumption
$\phi_{0,+}(x)=0$ is approximately
valid.

In Eq. (\ref{ap}), we have not included any diagonal elements in the 
coupling, compare with Eq. (\ref{q}). We neglect them in the condition
(\ref{neglectQ}) but in principle it would be also possible to 
absorb them by defining effective adiabatic potentials 
$\tilde{\lambda}_{\pm} = \lambda_{\pm} + mB^2/2$.  
\end{appendix}


\begin{thebibliography}{1}

\bibitem{ruschhaupt_2004_diode}
A. Ruschhaupt and J.G. Muga, {\rm Phys. Rev.} A {\bf 70}, 061604(R) (2004).

\bibitem{raizen.2005}
M.G. Raizen, A.M. Dudarev, Qian Niu, and N.J. Fisch,
{Phys. Rev. Lett.} {\bf 94}, 053003 (2005).

\bibitem{dudarev.2005}
A.M. Dudarev, M. Marder, Qian Niu, N.J. Fisch and M.G. Raizen,
{Europhysics Letters} to be published.

\bibitem{STIRAP}
K. Bergmann, H. Theuer, and B. W. Shore, 
Rev. Mod. Phys. {\bf 70}, 1003 (1998). 

\bibitem {schneble.2003}
D. Schneble, M. Hasuo, T. Anker, T. Pfau, and J. Mlynek,
{\rm J. Opt. Soc. Am. B} {\bf 20}, 648 (2003).

\bibitem{folman.2002}
R. Folman, P. Kr\"uger, J. Schmiedmayer, J. Denschlag, and C. Henkel,
{\rm Advances in Atomic, Molecular, and Optical Physics} {\bf 48}, 263 (2002).

\bibitem{HHM05} V. Hannstein, G. C. Hegerfeldt, J. G. Muga,
{J. Phys.} B: At. Mol. Opt. Phys. {\bf 38}, 409 (2005). 

\bibitem{singer.1982}
S. Singer, K.F. Freed, and Y.B. Band,
{\rm J. Chem. Phys.} {\bf 77}, 1942 (1982).

\bibitem{band.1994}
Y.B. Band and I. Tuvi,
{\rm J. Chem. Phys.} {\bf 100}, 8869 (1994).

\bibitem{ROS} B. Navarro, I. L. Egusquiza, J. G. Muga and G. C. Hegerfeldt, 
{Phys. Rev. A} {\bf 67}, 063819 (2003).  

\bibitem{messiah.book}
A. Messiah, {\it Quantum Mechanics} (Wiley, New York, 1958).



\end{thebibliography}
\end{document}